# Acoustic and Ultrasonographic Characterization of Neoprene, Beeswax, and Carbomer-Gel to Mimic Soft-tissue for Ultrasound

**AUTHOR NAMES AND AFFILIATIONS**

Debjani Phani (1), Rajasekhar K.V. (2), Anjali Thomas (3), Raghukumar Paramu (1), M. Suheshkumar Singh (3), Shaiju V.S. (1) , Venugopal Muraleedharan (1), R.K. Nair (1)

(1) Regional Cancer Centre, Thiruvananthapuram, Kerala, INDIA.

(2) Meenakshi Medical College Hospital and Research Institute, Chennai, Tamilnadu, INDIA

(3) Indian Institute of Science Education and Research Thiruvananthapuram, Kerala, INDIA.

**ABSTRACT:**

Materials with acoustic quantities similar to soft-tissue are essential as tissue-mimicking materials for diagnostic ultrasound (US). Acoustic quantity consists of the sound velocity ($c_{us}$), acoustic impedance (AI) and attenuation coefficient. In this work, the acoustic quantities of neoprene rubber, beeswax, and Carbomer-gel were determined. The $c_{us}$ and attenuation coefficient were estimated using the pulse-echo technique. The AI was calculated from the product of density and $c_{us}$. Results were compared with a benchmark based on the *International Commission on Radiation Units and Measurements Report-61*, *Tissue Substitutes, Phantoms and Computational Modelling in Medical Ultrasound*. The acceptance criteria were 1.043 ± 0.021 g/cm$^3$ (density), 1561 ± 31.22 m/s ($c_{us}$), 1.63 ± 0.065 MRayls (AI) and attenuation coefficients within 0.5-0.7 dB/cm/MHz. Computerized tomography (CT) and US images of specimens were obtained to compare with respective images of the human liver (a clinical soft-tissue), to evaluate the similarities in image contrast and echogenicity. Acoustic quantities of neoprene (density 1.45 g/cm$^3$, $c_{us}$ 1706 m/s, and AI 2.47 MRayls) and images were unacceptable. The density (0.96 g/cm$^3$) and CT images of beeswax were satisfactory. However, in beeswax the $c_{us}$ (2323 m/s), AI (2.23 MRayls) and its anechoic US images were unacceptable. Acoustic quantities of Carbomer-gel (density 1.03 g/cm$^3$, $c_{us}$ 1567 m/s, and AI 1.61 MRayls, and attenuation coefficient 0.6 dB/cm/MHz) were satisfactory. Carbomer-gel images could efficiently mimic the contrast and echogenicity of liver images. The uncertainties in $c_{us}$ measurements were 0.36 %, 0.11 % and 0.28 % for neoprene-50, beeswax and C-gel respectively. The attenuation coefficients had uncertainties 4.2 %, 1.9 %, and 2.6 % in these samples. The results of Carbomer-gel could resemble soft-tissue for US. It contains 95 % water, is effortless to prepare, and can support in developing a low-cost phantom for periodic performance evaluation of US scanners and contribute to patient care.

**1 INTRODUCTION**

An ultrasound (US) phantom is essential for quality control and periodic evaluation of diagnostic US equipment. These phantoms consist of tissue-mimicking materials (TMM) that can produce US images equivalent to a clinical setup. Commercial phantoms are highly expensive. Unfortunately, in many countries providing an US scanning facility to every patient is a problem and implementing a periodic quality assurance program along a phantom is a financial burden. This often results in disregarding quality assurance for US scanners. Indigenous phantoms mostly developed for clinical training, have used biodegradable TMMs like gelatine and agar [1–3]. However, for a quality assurance phantom, the TMM should have a longer shelf-life. Secondly, to mimic a tissue for US, this material should have acoustic quantities, similar to the tissue for the range of frequencies (~ 3 MHz – 25 MHz) used during US imaging. To explore low-cost, stable TMMs in this work, we aim to experimentally determine acoustic quantities of three different materials. The acoustic quantity is a set of physical parameters, most critical are the velocity of US ($c_{us}$), acoustic impedance (AI), attenuation coefficient, back-scattering coefficient, and the non-linearity parameter [4]. Ideally, a TMM should have the $c_{us}$ equal to 1540 m/s (± 10 m/s), AI of $1.6 \times 10^6$ Rayls and attenuation coefficient of 0.5 - 0.7 dB/cm/MHz, for frequencies ranging from 2 to 15 MHz, with a linear response of attenuation to frequency [5,6]. A material with similar values will ensure image resolution, contrast and depth of penetration equivalent to soft-tissue during US imaging. Moreover, the US images of these samples were also analysed for their ability to mimic the human liver (a homogenous soft-tissue).

The $c_{us}$, AI, and attenuation coefficient in neoprene rubber, beeswax, and carbomer-gel (*C-gel*) were determined. The present samples were chosen for their stability, low-cost, and/or the $c_{us}$ being close to 1540 m/s. ATS –Laboratories (Bridgeport, CT), presently CIRS Inc. (Norfolk, VA) uses urethane rubber ($c_{us}$ ~1450 m/s, desired > 5 %) as TMM in an US phantom. But, neoprene was selected in this study because of its reported $c_{us}$~ 1600 m/s (desired < 3.9 %) [7]. Furthermore, the hardness for a soft rubber is designated by its shore-A value; experimentally determined using the American Society for Testing and Materials (ASTM) Standard D-2240 [8]. The present sample of neoprene was prepared to have the shore-A value of 50 (neoprene-50). The second sample was a block of beeswax that was selected because of its stability. The third sample was a water-based gel. Water is universally available and it can transmit US signals. For increasing the viscosity of water carbomer-gel (*C-gel*) was prepared. This *C-gel* contained water (95 %) and carbomer-940 (5 %). Carbomers are synthetic high molecular weight polymers of acrylic acid (polyacrylic acid) commonly used in pharmacology [9,10]. Carbomer-940 can produce highly viscous stable gels that do not flow at normal room temperature.

Mass densities (density) of the prepared samples were determined from the computerized tomography (CT) scan and also using Archimedes Principle. The $c_{us}$ in the test samples was measured using the time of flight method and the AI was then calculated from the product of density and $c_{us}$. Finally, the attenuation coefficient (dB/MHz/cm) of each sample was measured using Beer-Lambert's Law. Furthermore, the CT and US scan images of each sample were compared with the respective images of the liver for similarity in image contrast and echogenicity. The results of the acoustic quantities were compared with the values given in Table 1.

TABLE 1 Benchmark to characterize the acoustic quantities of the samples

| *Density (g/cm³)* | *$c_{us}$(m/s)* | *AI (MRayls)* | *Attenuation Coefficient (dB/cm/MHz)* |
|---|---|---|---|
| *1.043* | *1561* | *1.63* | *0.5-0.7* |

Table 1: Source: Empirical relationships between acoustic parameters in human soft tissues, T.D. Mast (2000) derived from ICRU Report -61 and Duck F.A. (1990) [11-13].

Table 1 presents the average value of acoustic quantities for soft-tissues obtained from the International Commission on Radiation Units and Measurements (ICRU) Report 61 and a reference by Duck F. A. reported by Mast et al [11–13]. Here, the AI was obtained from the product of density and $c_{us}$. This table was used as a benchmark in the current work, to assess the acoustic quantities of the present samples. In an US image, the tissues of muscles, tendons, ligaments, fascia, fat, fibrous tissue, synovial membranes, nerves, and blood vessels are known as soft-tissues [14]. The reported reference values by T.D. Mast show the $c_{us,}$ density, and attenuation coefficients were different in these tissues [11]. The percentage variations of $c_{us}$ and density were 10 % and 16.3 % respectively in this data. Hence, it is practically reasonable to allow an acceptable range ±2 % for the acoustic quantities.

Based on these criteria in this study, the following conditions were set, to qualify a material as TMM for US.

i) The density should be 1.043 ± 0.021 g/cm³, $c_{us}$ 1561 ± 31.22 m/s and AI 1.63 ± 0.065 MRayls in a sample. These ranges allow the density and $c_{us}$ to vary within ± 2 % (refer to Table 1) and AI within ± 4 %. Attenuation coefficients should be in the range 0.5-0.7 dB/cm/MHz.

ii) The CT and US images of a sample compared to the respective scan of the normal human liver should resemble each other for image contrast and echogenicity, according to the recommendations of ICRU-61.

## 2 MATERIALS AND METHOD

The steps involved in determining the acoustic quantities of the samples are illustrated in Figure 1.

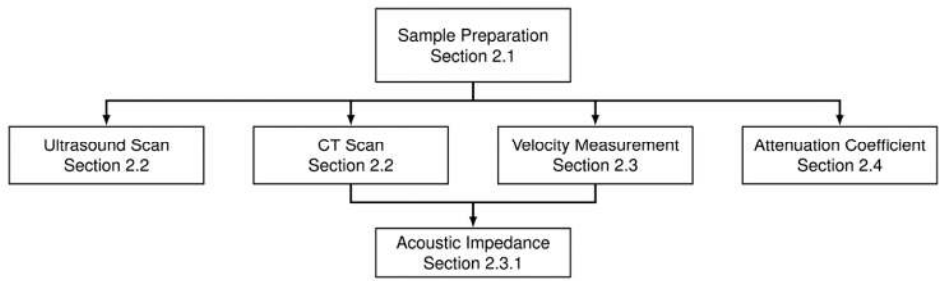

Fig.1. **Steps involved in determining the acoustic quantities -** Prepared samples were imaged in a computerized tomography (CT) and an ultrasound scanner. The velocity and attenuation coefficient of ultrasound were determined using the pulse-echo technique. The product of density (obtained from the CT) and velocity gave the acoustic impedance.

## 2.1 SAMPLE PREPARATION

### 2.1.1 NEOPRENE RUBBER

Neoprene-50 was compounded by a professional rubber products manufacturer. A cylindrical stainless-steel die was cast for the required dimension of neoprene (Diameter = 5 cm and height = 7 cm).

*Ingredients:* Raw neoprene (Skyprene® B-30), Magnesium Oxide (MgO), Antioxidant (AO45), Ultraflow (UF500), Carbon Black (HAF 330, 25 phr), Precipitated Silica, Naphthalene Oil (N-oil, Elasto 541), Zinc Oxide (ZnO), Ethylene Thiourea (Na22), and Tetra Methylthiuram Disulphide (TMTD).

*Mixing:* Neoprene rubber was masticated in a rubber mill for 10 min, MgO was added to the rolling band and mixed. UF500 and AO45 were added one at a time and mixed in between. To this HAF330, precipitated silica and N-oil were added across the mill and again mixed for 10 min. Finally, ZnO, NA22, and TMTD were added. The compound was mixed thoroughly to ensure a homogeneous mix. The dosage of carbon black (HAF 330) decides the shore value of rubber and antioxidants ensure good ageing properties and stability.

*Molding:* The rubber sample was molded in a hydraulic press. The prepared stainless-steel die was preheated, compounded mix was poured, and appropriate feed weight was placed onto it in a press of dimension 50.8× 55.9× 45.7 cm$^3$. After setting the curing parameters (Temperature= 423.15 K and Pressure = 15 MPa) the press was closed. The product was taken out after 30 min. Figure 2-a shows the prepared sample of neoprene.

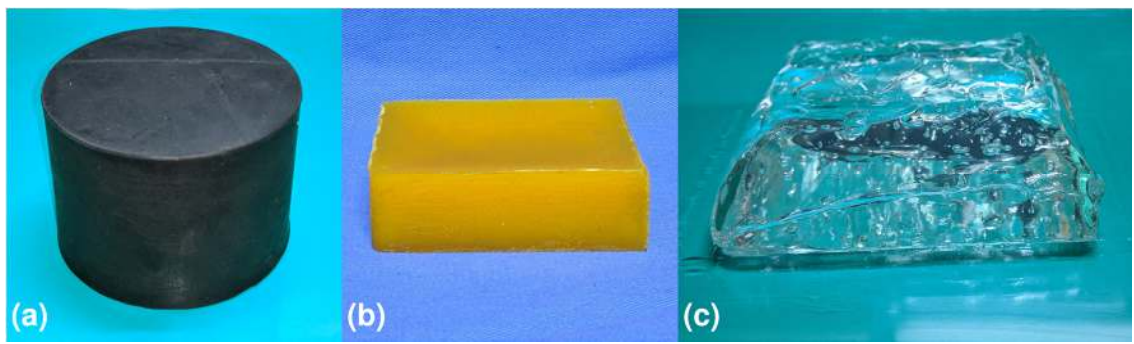

Fig.2. **Sample Photographs**: (a) the compounded neoprene (Shore-A value 50), (b) the block of beeswax, and(c) the prepared carbomer-gel at 26 °C.

*2.1.2 BEESWAX*

Filtered beeswax blocks (Hill Dews™) were purchased. A cavity (diameter ~ 3mm) was made in the wax. This was filled with water and wax sealed. The aim was to view this water-well in CT and US scans. The specimen of beeswax is shown in Figure 2-b.

*2.1.3 CARBOMER GEL*

***Ingredients:*** Distilled water, ethanol (Merck CAS# 64-17-5), carbomer-940 powder ( Carbogel® -940 Maruti Chemicals) and triethanolamine (Spectrum Chemicals).

***Procedure:*** 100 ml of an ethanol-water mixture (70 %) was taken in a beaker. While stirring on a magnetic stirrercarbomer-940 powder (5g, 5 % carbomer in the solution) was sifted into it. After thoroughly mixing, the solution was filtered into a jar preventing any lump. To these three drops of triethanolamine were added. The jar was closed and mixed vigorously forming a viscous gel. The prepared *C-gel* sample at 26° C is shown in Figure 2-c.

**2.2 COMPUTERIZED TOMOGRAPHY AND ULTRASOUND SCAN**

A 16-slice CT (Optima CT580 W) and US scanner (LOGIQ P5) from GE Healthcare were used to scan the samples. CT was performed to ascertain the samples were free from air-pockets and to obtain their densities. Exposure parameters of 120 kV, 277 mA, and a slice thickness of 0.63 mm were used during CT imaging. The CT DICOM (Digital Imaging and Communications in Medicine) images were then exported to the Eclipse™ *v*-13.6 (Varian Medical Systems, Palo Alto, CA) system. The Eclipse™ radiation therapy planning system was commissioned using a calibration curve of the Hounsfield Unit (from CT data) and material densities using a CT phantom (CATPHAN 500) with known in-homogeneities [15,16]. Eclipse™ could display the density on a CT image at any point of interest, using the *Physical Parameters Tool* in the *External Beam Planning* Application. The uncertainty in density measurements in the Eclipse™ system (± 0.36 %) was determined from the deviation of the average density at 10 points in a region of interest in the CATPHAN image and the actual density of this region from the CATPHAN data-sheet.

US scans were performed on the present samples using a 4C curvilinear probe (nominal frequency 4 MHz), having a footprint of 18 × 66 mm, bandwidth 1.6 - 4.6 MHz and the gain ranging from 76-92 during imaging. The US scanning was carried out and evaluated by an experienced radiologist to establish the ability of a sample to resemble a soft-tissue background and the liver scans.

## 2.3 ULTRASOUND VELOCITY MEASUREMENT

The experimental set-up used for measuring the $c_{us}$ and the attenuation coefficient is illustrated in Fig.3-a. The experimental system used for the experiments shown in Fig. 3-b, composed of a pulser-receiver (5073PR-40-P, Olympus, with pulse repetition frequency of 100 Hz, 25MHz nominal frequency, and pulse width ~ nano sec), a focusing US transducer (V 375 - SU, Olympus with 19 mm focal length, focal spot size ~ 153 µm, and focal zone length of 4 mm), a 3-axis step-by-step motorized positioning system (Newmark NSC-G Series, Newmark Systems Inc.) and the data acquisition (DAQ) system (779745-02, NI PCI-5114). In time-sharing mode, the pulser-receiver also served as a sensor to receive electrical signals, corresponding to reflected acoustic pulses that could amplify weak signals (amplification ~ 39 dB). During the experiments, samples were mounted on an acoustic reflector (a stainless-steel plate) and fully immersed in a water bath (18 °C) along with the transducer surface for acoustic coupling. Samples were kept at the focal zone of the transducer for delivering maximum energy of acoustic signals and a train of acoustic pulses were delivered to the material. The acoustic echoes (A-line data) reflected from sample surfaces, were acquired by the DAQ system, after amplification, at a sampling frequency of ~ 250 MHz. Fig.3(c)-(e) shows the photographs of small sample sections of neoprene-50 ($10\times15\times1.93$ mm$^3$ obtained by slicing the sample in Fig.2-a using a slow motorized saw), beeswax ($20\times40\times2.64$ mm$^3$, prepared by melting the beeswax and moulding in a stainless-steel mould), and the C-gel ($18\times18\times4$ mm$^3$, by taking a small quantity inside a small glass container with open ends).

*Experimental Procedure:* To estimate the $c_{us}$, pulse-echo method was implemented. Incident US pulses reflected from the topmost (at time $t_1$) and the distal surface of the sample were obtained (time = $t_2$). The A-line data recorded in the experimental system for neoprene-50 is depicted in Fig. 4. Data were post-processed and the $c_{us}$ were measured using the time-of-flight equation. The time interval $\Delta t$ ($\Delta t = t_2 - t_1$) was the time taken by the acoustic pulses to travel two times across the thickness (d) of the sample. So, in time $\Delta t$, the pulses traverse the sample with an effective thickness $d_{eff}$ ($d_{eff} = 2d$).

Using the expression, $d_{eff} = c_{us} \times \Delta t$, the $c_{us}$ of the samples were obtained using the equation

$c_{us} = d_{eff}/\Delta t$

For each sample, the experiment was repeated ten times and data were averaged.

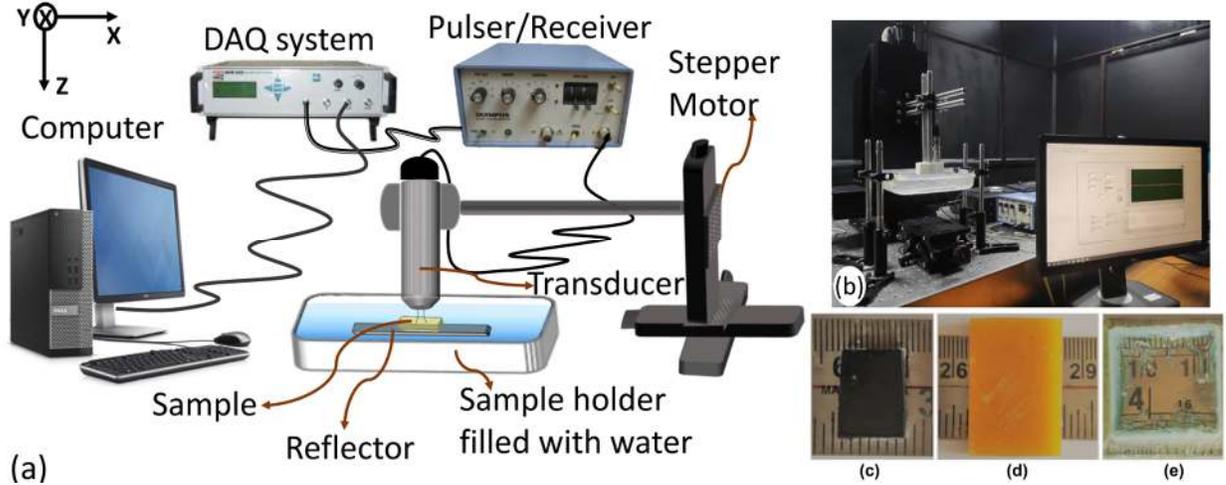

Fig.3(a) Schematic representation of the experimental set-up for measuring ultrasound velocity and the attenuation coefficient in a specimen. (b) Photograph of the experimental set-up. (c)-(e) Photographs of sample sections neoprene-50 (c), beeswax (d), and *C-gel* (e).

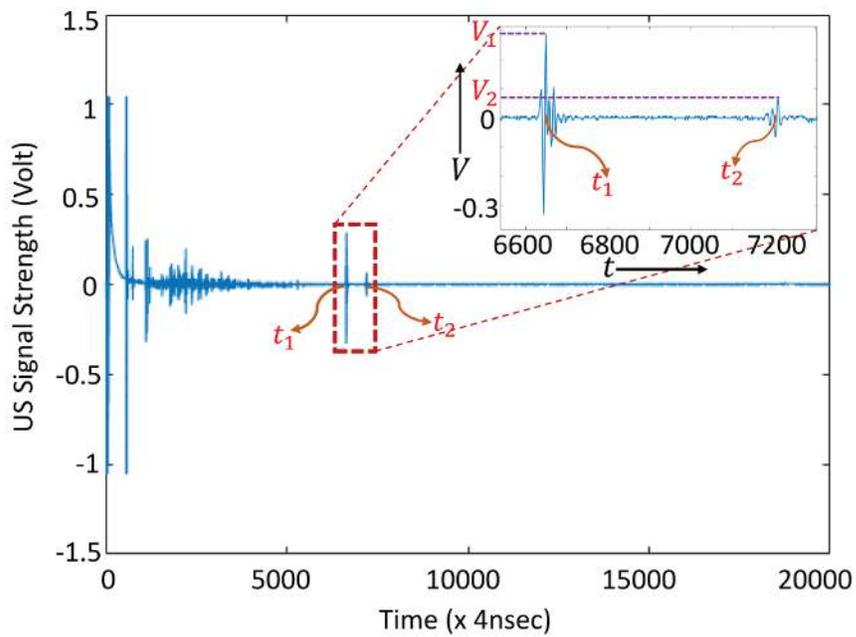

Fig.4. A-line plot of the neoprene-50 sample (thickness ~ 1.93 mm). The figure shows the reflected signals from the proximal ($V_1$) and distal ($V_2$) surfaces of the specimen, at times $t_1$ and $t_2$ respectively. The inset shows an enlarged view of the data within the dotted rectangular box.

## 2.3.1 ESTIMATION OF ACOUSTIC IMPEDANCE

The product of the density of a sample (from the Eclipse™) and experimentally determined $c_{us}$ in it gave the AI (kg/m²/s or Rayls). This was obtained for all the samples.

## 2.4 ATTENUATION COEFFICIENT MEASUREMENT

The acoustic attenuation coefficients for the present samples were measured using the set-up shown in Fig.3-a. Propagation of US waves in a material obeys Beer-Lambert's Law [17]. This law states that the intensity of acoustic waves decays exponentially while propagating through a material.

$$\text{Mathematically, } I = I_0 \, e^{-\mu \cdot d_{eff}}$$

Here, $I_0$ and $I$ are the intensity of the incident wave and the intensity after traversing through an effective distance $d_{eff}$ ($d_{eff} = 2d$) in the sample respectively and $\mu$ denotes the attenuation coefficient.

The strength of acoustic signals corresponding to pulses reflected from sample surface ($V_1$) and reflector ($V_2$), as shown in Fig.4 were measured using the maximum values of the Hilbert transform of the signal and subsequently, we could estimate $\mu$ (dB/cm) using

$$\mu = \frac{10}{2d} \log \frac{V_1}{V_2} \quad \text{dB/cm}$$

This value was divided by the transducer frequency (25 MHz) to obtain the attenuation coefficient in dB/cm/MHz.

## 3 RESULTS

Figure 5 presents the CT and US images of the prepared samples and of the human liver.

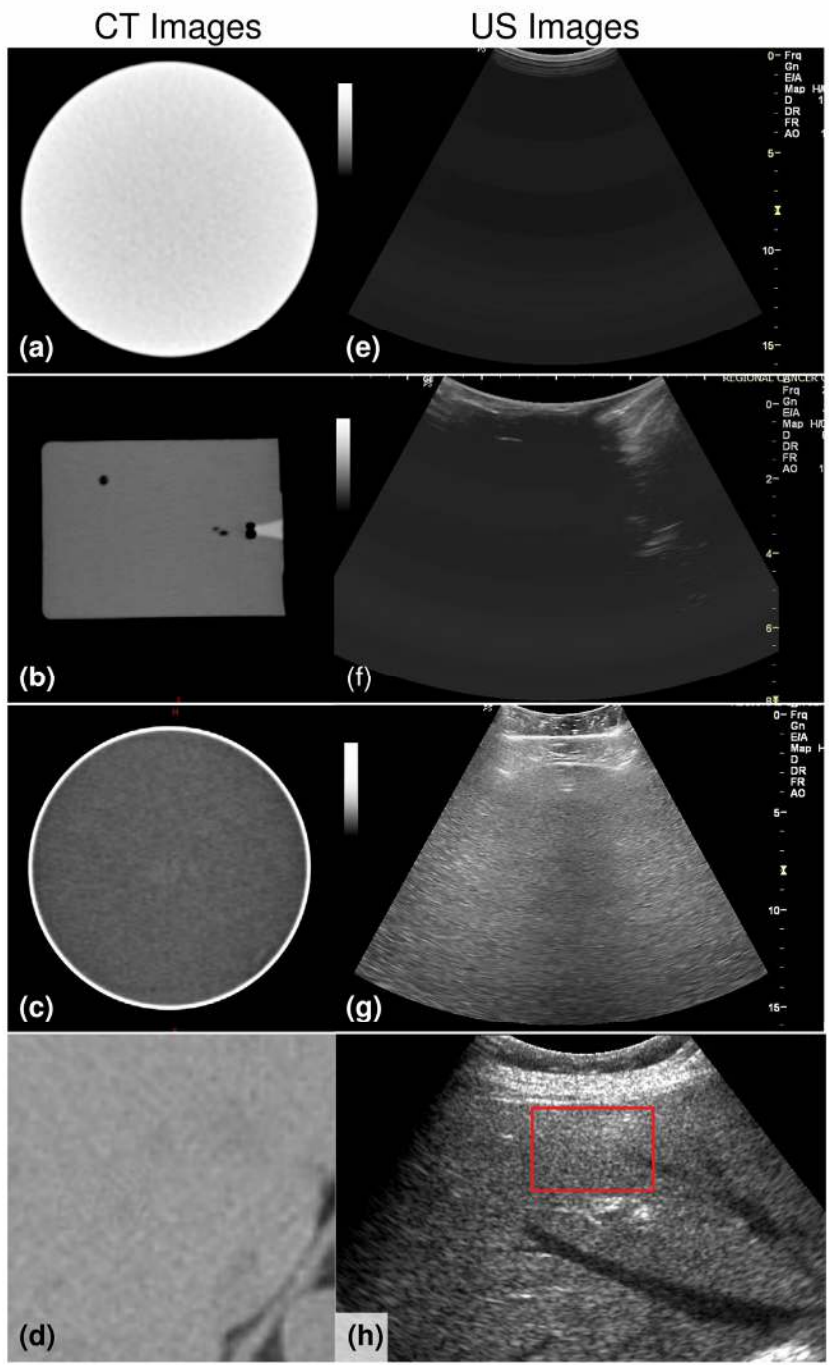

Fig.5. **Computerized Tomography and Ultrasound Images of the Samples**. (a) - (c) shows the CT images of neoprene-50, beeswax and the carbomer-gel respectively. (e) – (g) shows their respective US images. (d) and (h) are the CT and US of human liver, presented here as benchmark images representing soft-tissue (the box shows the region of interest to compare the US images).

Figure (5-a) – (5-c) shows the CT images of the neoprene-50, beeswax, and *C-gel*. The respective US images are shown in Figure (5-e) – (5-g). These images are compared to the benchmark CT and US scans of human liver shown in Figure 5-d and 5-h. The images (CT and US) of neoprene-50 are different from the liver images. Although the CT images of beeswax and liver are similar, their US images are not alike. However, it is observed that these images of *C-gel* and liver have similar image contrast and echo-texture.

Table 2 presents the densities of the samples measured from the Eclipse™ system and the physically measured densities. This table also indicates whether a material resembles soft-tissue/liver in the CT and US images.

Table 2. Results from the CT and US scans of the samples

| Sample | Density (g/cm$^3$) | | Acceptable (Yes/No) | |
|---|---|---|---|---|
| | Ratio† | Eclipse™ | CT-Image | US-Image |
| Neoprene-50 | 1.38 | 1.45 | No | No |
| Beeswax | 0.98 | 0.96 | Yes | No |
| C-gel | 1.01 | 1.03 | Yes | Yes |

Table 2. Sample densities from the ratio and Eclipse™ system are produced. This table mentions if an image resembles the corresponding liver image.
† Physically measured ratio of mass to volume of the samples using Archimedes Principle.

The density of neoprene-50 (1.45 g/cm$^3$) is the highest among the present samples followed by *C-gel* (1.01 g/cm$^3$) and beeswax (0.98 g/cm$^3$). The *C-gel* images resemble the liver images therefore; the image results of *C-gel* are acceptable.

Table 3 lists the experimentally determined acoustic quantities for the present samples. This table gives the percentage variation of each parameter compared to the corresponding value in benchmark Table 1.

Table 3. Experimental results of acoustic quantities in the samples and their variation compared to the referenced Table 1.

| Sample | Density (g/cm³) | | $c_{us}$(m/s) | | AI (MRayls) | | Attenuation Coefficient (dB/MHz/cm) |
|---|---|---|---|---|---|---|---|
| | *Observed value* | *% variation* | *Observed value* | *% variation* | *Observed Value* | *% variation* | *Observed Value* |
| *Neoprene-50* | 1.45 | 39 | 1706 ± 6.2 | 9.29 | 2.47 | 51.5 | 0.69±0.029 |
| *Beeswax* | 0.96 | -7.96 | 2323 ± 2.5 | 48.81 | 2.23 | 36.8 | 0.51±0.010 |
| *C-gel* | 1.03 | - 1.25 | 1567 ± 4.4 | 0.38 | 1.61 | -1.23 | 0.60±0.016 |

Table 3. The measured density (Eclipse™ system), the velocity of ultrasound ($c_{us}$), acoustic impedance (AI), and attenuation coefficient are presented for the characterization of the acoustic properties of the samples. The percentage variation of these results from the values in Table 1 is given. The results of the $c_{us}$ and attenuation coefficient are averages of ten measurements. The standard deviation of measurement (for ten data sets) is mentioned. A negative value is obtained for results, lower than the benchmark value.

The uncertainty in $c_{us}$ measurements calculated using the values given in Table 3 is 0.36 %, 0.11 % and 0.28 % for neoprene-50, beeswax and C-gel respectively. Similarly, the uncertainties in the attenuation coefficient measurements are 4.2 %, 1.9 % and 2.6 % in the respective samples.

## 4 DISCUSSIONS

The present study was based on reported ICRU report-61 to determine the acoustic properties of three samples. Their images (CT and US) were compared with the liver images to evaluate their similarity.

### 4.1 NEOPRENE

From Tables 2 and 3 the density of neoprene-50 (1.45 g/cm³) is higher (desirable < 39 %) and unacceptable. This is also confirmed from Figure 5-a, where the CT image of neoprene resembles a denser material than the human liver (see Figure 5-d). A solution to improve the density can be obtained by comparing this sample with previous studies using neoprene [7,18]. Furthermore, the US image of neoprene is anechoic, as shown in Figure (5-e). Scattering is critical for producing echo details in B-mode images. This acoustic scattering caused by the in-homogeneity in a medium is a result of multiple reflections in a small volume from the changes in local $c_{us}$ and/or density. Absence of scattering prevents image formation during an US procedure. Similarly, in the case of neoprene, the anechoic image can be a consequence of the material lacking sources of acoustic scattering.

Table 3 shows the $c_{us}$ in neoprene 1706 m/s is higher by 9.3 % from the reference value 1561 m/s. Garu *et al* found the $c_{us}$ in neoprene -50 to be 1620 m/s in their samples. Our result for $c_{us}$ in neoprene-50 is higher by 86 m/s (5.3 %, shown in Table 3) from this value. It has been reported that the type of carbon black used for compounding neoprene significantly decides its acoustic quantities. Following the recipe and procedure meticulously is also crucial, as these results are sensitive to the compounding process [18]. However, the AI 2.47 MRayls is higher by 51.5 %. This large deviation in AI is primarily caused by the variation in density (39 %) from the desired value 1.043 g/cm$^3$ (refer to Table 1). However, the attenuation coefficient of neoprene 0.69 dB/cm/MHz falls within the desirable range 0.5-0.7 dB/cm/MHz and is acceptable. There is a significant deviation in the CT and US images of neoprene and liver in Fig.5. Although the attenuation coefficient is acceptable, the density, $c_{us}$ and AI have a deviation larger than our acceptance criteria mentioned in

1 INTRODUCTION. As the required conditions are not met the neoprene-50 sample does not qualify as a TMM for US. Although the results of the present rubber sample are not satisfactory, it is a desirable system given its stability, reproducibility and low-cost.

**4.2 BEESWAX**

From Table 2, the density of the beeswax 0.964 g/cc has a variation of 7.96 % (refer to Table 3) when compared to the reference value in Table 1. The CT image of the beeswax shown in Figure 4-b confirms that it is free from air-pockets and mimics the liver CT image in Figure 4-d. Although the water-well is visible in the CT scan, the sample is anechoic in the US images as shown in Figure (5-f). As pointed out in the case of neoprene, the scattering in beeswax is inadequate to form an US image. The measured $c_{us}$ in beeswax (2323 m/s) is higher than our requirement and the calculated AI given in Table 3, is higher by 36.8 % from the benchmark. The higher AI is a result of higher $c_{us}$ as the density of beeswax is acceptable.

The attenuation coefficient of US in beeswax 0.51 dB/cm/MHz is acceptable. However, the present beeswax could not resemble the liver in the US images and the acoustic quantities have a large variation. Since, the required conditions of imaging and acoustic quantities are not satisfied, we find beeswax unsuitable for the intended application.

Waxes are low-cost, stable, and simple systems that can be easily moulded into desired shapes. In an attempt to reduce $c_{us}$ in wax, investigators have explored materials like paraffin-gel wax, olefin polymers in mineral oil, and polyurethane gel for their tissue-mimicking properties for US [19–21]. In future work, the gel-wax system will be evaluated for similar applications.

**4.3 CARBOMER GEL**

The density (1.03 g/cm³) of the *C-gel* compared to the reference value in Table 1, is acceptable (variation, 1.3 % refer to Table 3). In Figure 5-c, the CT image of *C-gel* shows a homogeneous sample that is similar to the liver CT in Figure 4-d. The US images of the gel (shown in Figure 5-g) is echogenic with echogenicity similar to the US image of liver (refer to Figure 5-h). The resultant $c_{us}$ (1567 m/s) and the calculated AI (1.61 MRayls) vary by 0.38 % and 1.2 % respectively in *C-gel*, refer to Table 3. Its attenuation coefficient 0.6 dB/cm/MHz is within the acceptable range. These results of *C-gel* satisfy the criteria of images and the maximum variation of any acoustic quantity is 1.2 % (for AI). Therefore, the present findings suggest that the *C-gel* can represent soft-tissue for US. The surface of a phantom prepared using *C-gel* will not be firm to hold a transducer in place during a scan and this problem shall be addressed. The topmost surface made of gel-wax in a *C-gel* phantom can be attempted. A quality assurance phantom with a firm surface can have *C-gel* filled body to represent a clinical soft-tissue. A small prototype phantom using *C-gel* and gel-wax was tested and this was found to generate images similar to soft-tissue scans.

The present results establish that the *C-gel* represents a uniform soft-tissue background in the CT and US images. These scans show contrast and echo-texture similar to the liver images; refer to the last two images in both the columns of Figure 4. The present gel prepared at 26 °C can hold its shape and is non-drip. Moreover, this low-cost gel is stable under normal room conditions, free from microbial infestation, and is effortless to prepare. Thus, the *C-gel* can provide a practical solution in making a low-cost TMM for the US that can be renewed effortlessly, in any clinical set-up, and hence can be widely employed in similar applications.

**5 CONCLUSIONS**

In the present study acoustic and ultrasonographic characterization of neoprene-50, beeswax and *C-gel* systems were performed to determine their ability to mimic soft-tissue for US. This required measuring their acoustic quantities (a set of parameters). The results were compared with benchmark values based on ICRU Report-61. Secondly, the CT and US images of these materials were compared to the contrast and echo-texture of human liver scans. For each test sample, the experimental results supported their CT and US scans. The acoustic quantities of neoprene-50 substantially varied from the reference values and its US images were anechoic. A rubber-based US phantom can last a lifetime with proper handling; therefore, a rubber having acceptable results remains a desirable material. Although the density and CT image of the beeswax sample were satisfactory, its US images were anechoic. The CT and US images of *C-gel* resembled the liver scans and thus, represented a soft-tissue background. The density, $c_{us}$, and AI in *C-gel* were very close to the reference values. These results demonstrate *C-gel* to achieve the requirements better than any other sample in this study.

Although the temporal stability of the present gel is not studied here, commercial carbomer-940 gels are known to have a shelf-life of two to three years. *C-gel* is a non-toxic system that is effortless to prepare. Containing 95 % water this gel has a simple formulation that can be reproduced widely. These attributes make this system highly desirable. The acoustic properties of *C-gel* have not been studied during the reporting of this work. From the present findings, we suggest *C-gel* can be used as a TMM for US. An US phantom based on *C-gel* can offer an alternative to costly phantoms that will allow diagnostic US facilities to incorporate a periodic quality assurance program. This will help in a consistent, reliable and error-free reporting of this highly popular diagnostic tool.